\documentclass[a4paper]{article}

\usepackage{INTERSPEECH2019}
\usepackage{multirow}
\usepackage{lipsum,url}
\usepackage{hhline}
\usepackage{caption} 
\usepackage{hyperref}
\usepackage{tabu}
\usepackage{xcolor}
\usepackage{verbatim}

\def\vec#1{\ensuremath{\bm{{#1}}}}

\title{I4U Submission to NIST SRE 2018:\\Leveraging from a Decade of Shared Experiences}

\name{Kong Aik Lee$^1$, Ville Hautam\"{a}ki$^2$, Tomi Kinnunen$^2$, Hitoshi Yamamoto$^1$, Koji Okabe$^1$, \\ Ville Vestman$^{1,2}$, Jing Huang$^3$, Guohong Ding$^3$, Hanwu Sun$^4$, Anthony Larcher$^5$, Rohan Kumar Das$^6$, Haizhou Li$^6$, Mickael Rouvier$^7$, Pierre-Michel Bousquet$^7$, Wei Rao$^8$, Qing Wang$^{9}$, Chunlei Zhang$^{10}$, Fahimeh Bahmaninezhad$^{10}$, Hector Delgado$^{11}$, Jose Patino$^{11}$, Qiongqiong Wang$^1$, Ling Guo$^1$, Takafumi Koshinaka$^1$, Jiacen Zhang$^1$, Koichi Shinoda$^1$, Trung Ngo Trong$^2$, Md Sahidullah$^2$, Fan Lu$^3$, Yun Tang$^3$, Ming Tu$^3$, Kah Kuan Teh$^4$, Huy Dat Tran$^4$, Kuruvachan K. George$^4$, Ivan Kukanov$^4$, Florent Desnous$^5$,  Jichen Yang$^6$, Emre Y{\i}lmaz$^6$, Longting Xu$^6$, Jean-Francois Bonastre$^7$, Chenglin Xu$^8$, Zhi Hao Lim$^8$, Eng Siong Chng$^8$, Shivesh Ranjan$^{10}$, John H.L. Hansen$^{10}$, Massimiliano Todisco$^{11}$, and Nicholas Evans$^{11}$}
\address{
  $^1$NEC Corporation :: Tokyo Institute of Technology, Japan \\
  $^2$University of Eastern Finland, Finland :: INRIA, France \\
  $^3$JD AI Research and Platform, USA -- $^4$Institute for Infocomm Research, Singapore\\
  $^5$LIUM, France -- $^6$National University of Singapore, Singapore -- $^7$LIA, France \\
  $^8$Nanyang Technological University, Singapore -- $^9$Northwest Polytechnic University, China\\
  $^{10}$CRSS, University of Texas at Dallas, USA -- $^{11}$EURECOM, France}
\email{k-lee@ax.jp.nec.com, villeh@cs.uef.fi, tomi.kinnunen@uef.fi}
\begin{document}

\maketitle
\begin{abstract}
The I4U consortium was established to facilitate a joint entry to NIST speaker recognition evaluations (SRE). The latest edition of such joint submission was in SRE 2018, in which the I4U submission was among the best-performing systems. SRE'18 also marks the 10-year anniversary of I4U consortium into NIST SRE series of evaluation. The primary objective of the current paper is to summarize the results and lessons learned based on the twelve sub-systems and their fusion submitted to SRE'18. It is also our intention to present a shared view on the advancements, progresses, and major paradigm shifts that we have witnessed as an SRE participant in the past decade from SRE'08 to SRE'18. In this regard, we have seen, among others, a paradigm shift from supervector representation to deep speaker embedding, and a switch of research challenge from channel compensation to domain adaptation.
\end{abstract}

\section{Introduction}
The series of speaker recognition evaluations (SRE) conducted by NIST has been a major driving force advancing speaker recognition technology~\cite{MARTIN2000,Martin2007}. The basic task is speaker verification: given a segment of speech, decide whether a specified target speaker is speaking in that segment. The SRE in 2018 marks the most recent and ambitious attempt to tackle more realistic tasks~\cite{sre18}. 

The SRE'18 evaluation set comprises two partitions -- \emph{Call-My-Net 2} (CMN2) and \emph{Video-Annotation-for-Speech-Technology} (VAST) -- named after the corpora \cite{Jones2017,Tracey2018} from which the data were derived. For the CMN2 partition, domain mismatch appears to be the major challenge -- the \emph{train set} consists of English utterances while the \emph{test set} consists of Tunisian Arabic utterances. For the VAST partition, the major challenge is the \emph{multi-speaker test} scenario, for which an additional diarization module has to be used to determine the target speaker (if any) from a given test segment. This paper presents the technical details of the datasets, sub-system development, and fusion strategy of I4U SRE'18 submission.

In the past decade, I4U participated in five SREs, namely SRE'08, 10, 12, 16, and 18 \cite{hli2009,hli2010,Saeidi2013,Lee2017,i4u2018}. Aside from a joint submission, the I4U consortium was formed with a common vision to promote research collaboration and facilitate active exchange of information and experience towards the open evaluation of speaker recognition technology. Along the way we have seen \emph{old} technical challenges were solved, \textit{e.g.}, channel compensation~\cite{nap,Kenny2007,kenney2010plda}, after which researchers have moved on to tackle new challenges, \textit{e.g.}, domain adaptation~\cite{Romero2014,alam2018coral,kalee2018,Sun2016}. SRE’18 marks the ten-year anniversary of I4U consortium into NIST SRE series of evaluation. As we set out with the aim to tackle new frontiers in robust speaker recognition, we reckon that it is beneficial looking into past I4U submissions, to share the lessons learned and the insights gained from a decade of I4U experiences. 

The paper is organized as follows. Section 2 gives a brief description on SRE'18 dataset, the challenges, and the I4U solutions to deal with them. Then, we present the I4U SRE'18 results in Section 3. Section 4 looks into past I4U submissions. Section 5 concludes the paper.

\section{Data and Challenges}
\label{sec:train_and_dev}
Two main challenges of SRE'18 are (i) \textbf{domain mismatch} in the CMN2, and (ii) \textbf{multi-speaker test} segment in VAST. In this section, we provide a brief description on the CMN2 and VAST data conditions that give rise to the aforementioned challenges and elaborate on the strategy and techniques implemented in I4U sub-systems to deal with them.         

\subsection{CMN2 and VAST Partitions}
Table \ref{table:dataset} shows the list of corpora made available for the fixed-training condition of SRE'18. The train, development and evaluation sets consists of two partitions~\cite{sre18}, namely, the \emph{Call-My-Net 2} (CMN2)~\cite{Jones2017} and \emph{Video-Annotation-for-Speech-Technology} (VAST)~\cite{Tracey2018}. 
\begin{itemize}
    \item \textbf{CMN2} partition comprises conversational speech in Tunisian Arabic recorded over \emph{voice over internet protocol} (VOIP), in addition to the \emph{public switch telephone network} (PSTN). This is different from Fisher, Switchboard and the Mixer corpora used in previous SREs. Comparing \texttt{CMN2-Train} to the \texttt{CMN2-Dev} and \texttt{CMN2-Eval} sets (see Table~\ref{table:dataset}), two major differences are languages (English versus Tunisian Arabic) and transmission channels (a mix of VOIP and PSTN versus PSTN only). These differences lead to the so-called \textbf{domain mismatch} problem, in which the test set does not follow the same distribution as the train.
    
    \item \textbf{VAST} partition comprises wideband English speech segments extracted from amateur video recordings downloaded from YouTube\textsuperscript{\textregistered}. A signature feature of the VAST partition is multi-speaker conversation with considerable background noise. The VoxCeleb~\cite{Nagrani2017} and SITW~\cite{McLaren2016} used as the \texttt{VAST-Train} and \texttt{VAST-Dev}, as shown in Table~\ref{table:dataset}, bear the same properties and therefore the same domain. 
\end{itemize}

\vspace{1ex}
\noindent While it might seem unusual to include two distinct data partitions in a single core task, the setup enables a systematic comparison to past results and system performance on new tasks. In this regard, the CMN2 partition is the continuation of past SREs with new challenges (domain mismatch and lack of labelled in-domain data), while the VAST partition represents a new initiative towards speaker recognition in the wild. See Figure~\ref{fig:sres}. We shall touch upon this point further in Section~\ref{sec:pastandfuture}.

\begin{table}[t]
	\caption{List of speech corpora designated as train and development sets for SRE'18 CMN2 and VAST~\cite{sre18}.}
	\vspace{-4ex}
	\label{table:dataset}
	\footnotesize
	\begin{center}
	\begin{tabular}{l l l}
		\noalign{\hrule height 0.75pt}
		\textbf{Partition} & \textbf{Corpus} & \textbf{Language}\\
		\hline \hline
		\texttt{CMN2-Train}        & SRE'04-05-06-08-10-12	                & \multirow{4}{*}{English (PSTN)} \\
							  	    & Swb-2 Phase I, II, III 	            & \\
								    & Swb-Cell Part 1, 2 	                & \\
								    & Fisher 1, 2 			                & \\  
		\texttt{CMN2-Dev}           & SRE'18-Dev                            & Tunisian Arabic \\
		                            & SRE'18-CMN2-Unlabeled                 & (PSTN + VOIP)   \\
		\texttt{CMN2-Eval}          & SRE'18-Eval                           & \\                          
		\hline 
		\texttt{VAST-Train}         & VoxCeleb1, VoxCeleb2                  & \multirow{3}{*}{English (wideband)} \\
		\texttt{VAST-Dev}           & SRE'18-Dev, SITW-Eval                 & \\
		\texttt{VAST-Eval}          & SRE'18-Eval                           & \\
		\noalign{\hrule height 0.75pt}
	\end{tabular}
	\end{center}
\end{table}

\subsection{Domain adaptation}
\label{sec:domain_adaptation}

A state-of-the-art speaker recognition system consists of a speaker embedding front-end (\emph{e.g.}, i-vector~\cite{Dehak10frontend}, x-vector~\cite{snyder2018vector}), followed by a scoring back-end, which is typically implemented with the \emph{probabilistic linear discriminant analysis} (PLDA) \cite{ioffe2006plda, Princepaper}. One advantage of the two-stage pipeline is that the same feature extraction and speaker embedding front-end could be used while domain adaptation is accomplished via a transformation on the x-vectors (or i-vectors)~\cite{Sun2016,alam2018coral}, or the parameters of the PLDA model~\cite{kalee2018}, to cater for the condition in the anticipated application. The two-stage pipeline design was used for all the twelve sub-systems in I4U SRE'18 submission. 

In the case of CMN2, the speaker embedding front-end and PLDA backend are trained on the \textbf{out-of-domain} \texttt{CMN2-Train} dataset. Let $\boldsymbol{\phi}$ be the speaker embeddings (\emph{i.e.}, x-vector or i-vector). A PLDA model is given by 
\begin{equation}
p\left(\boldsymbol{\phi}\right) = \mathcal{N}\left(\left. \boldsymbol{\phi} \right| \boldsymbol{\mu}, \mathbf{\Phi}_{\rm b}+\mathbf{\Phi}_{\rm w} \right),
\notag
\end{equation}
where $\boldsymbol{\mu}$ is the global mean,  $\boldsymbol{\Phi}_\text{b}$ and $\boldsymbol{\Phi}_\text{w}$ are the between and within-speaker covariance matrices of full rank, respectively. Given \texttt{SRE’18-CMN2-Unlabeled}, an unlabelled set of \textbf{in-domain} data (see Table~\ref{table:dataset}), the central idea of domain adaptation is to estimate the in-domain between and within-speaker covariance matrices from the in-domain, yet unlabelled, dataset with some helps from out-of-domain covariance matrices. In I4U SRE'18 submission, two unsupervised domain adaptation techniques have been found to be useful, namely, (i) model-level \emph{correlation alignment} with CORAL+~\cite{kalee2018}, and (ii) Kaldi's PLDA adaptation\footnote{https://github.com/kaldi-asr/kaldi/tree/master/egs/sre16/v2}. We refer the interested reader to \cite{alam2018coral,kalee2018,Sun2016} and references therein for more details.


\subsection{Multi-speaker test segment}
\label{sec:multi-speaker}
The multi-speaker test scenario is not new. It first appeared in NIST SRE'99 \cite{MARTIN2000} where a summed two-channel telephone speech consisting of two speakers was used as the test segment. For the case of SRE'18 VAST partition, there may be several speakers in a test segment. One straightforward solution is to score the entire test segment regardless of other competing speakers. Alternatively, one could use a diarization system to obtain several speaker clusters, score the enrollment segment against all the speaker clusters and select the maximum score. Speaker diarization was explored in Sys. 6 and 7 as shown in Table~\ref{table:subsysperf}

Following \cite{shell2014diarization}, speaker diarization was accomplished using an x-vector PLDA system. Given a VAST test segment, it is first split uniformly into cuts of about 1 second, which are then represented as x-vectors. A matrix of PLDA scores is computed from all the cross-pairs of these x-vectors. The score matrix is used as the affinity matrix in \emph{hierarchical agglomerative clustering} (AHC) where speaker clusters are derived. The number of clusters is determined by an AHC stopping threshold tuned on the \texttt{SITW} set. It is worth mentioning that speaker change point detection which has shown to be critical in reducing the diarization rate seem to be less important in reducing the error rate in speaker verification task. 

\section{I4U SRE'18 Submission and Results}
The sub-system performance is shown in Table~\ref{table:subsysperf}. Among the twelve sub-systems, eight of them employed x-vector embedding in some form. Notably, Sys. 5 and 6 use attentive pooling layer in the x-vector extractor, while Sys. 10 uses a t-vector embedding trained with a triplet loss~\cite{zhang2018text}. The remaining three sub-systems use i-vector. Comparing the results, x-vector gives a much better performance than i-vector on both CMN2 and VAST. The Kaldi PLDA domain adaptation was the most commonly used strategy. The CORAL+ was also successfully employed resulting in the lowest EER and $C_{\mathrm{prim}}$. Clustering unlabeled set to obtain pseudo-speaker labels was tried in Sys. 3, though no significant difference between clustering and Kaldi adaptation strategy is observed. In terms of the performance on the VAST partition, we observe only slight benefit in using speaker diarization (Sys. 6 and 7) suggesting a good potential for further improvement.   

The scores of the sub-systems were pre-calibrated before fusion. To this end, we apply an affine transformation with simple scaling factor and bias to the scores. The calibrated scores from sub-systems were then combined with a linear fusion. 
The cross-entropy cost was used for the calibration and fusion with a slight different setting on the effective prior. In this regard, the effective prior was set to 0.5 for score calibration, while an effective prior $P_{\mathrm{eff}}$ of 0.005 and 0.05 was used for the fusion for CMN2 and VAST partitions, respectively. Note that the effective priors were set based on those specified in the evaluation plan~\cite{sre18}. The BOSARIS Toolkit \cite{brummer} was used to perform calibration and fusion. In the primary submission, only subsystems with positive weights were retained. This resulted in 7 subsystems in primary submission of the CMN2 partition (Sys. 3, 4, 6, 7, 9, 10 ,11), and 11 subsystems in the primary submission of the VAST partition (Sys. 1 to 11). 

The final submitted fusion system performance is shown in Table~\ref{table:performance}. In general, the performances on development set and evaluation set agree on the CMN2 partition. On the VAST partition, we notice a large performance gap between development and evaluation sets where the EER increases from 3.70\% to 10.18 \%. This result reflects the lack of suitable development set for the VAST data. This justifies the use of \texttt{SITW} as \texttt{VAST-Dev} as shown in Table~\ref{table:dataset}.

\renewcommand{\arraystretch}{1.0}
  \begin{table}[t] 
  	\caption{Sub-system performance on the NIST SRE'18 evaluation set. Performance is measured in terms of EER and min $C_{\mathrm{prim}}$. We indicate, whether VAST system used diarization (2 systems) and what type of domain adaptation (DA) was utilized (Kaldi PLDA adaptation, CORAL+ or obtaining pseudo labels from the unlabeled set by clustering). Tag `\texttt{i}' indicates an i-vector system, tag `\texttt{t}' indicates a t-vector, tag `\texttt{x}' indicates an x-vector, while tag `\texttt{x+}' indicates an x-vector with attentive pooling.} 
  	\vspace{-4ex}
  	\begin{center}
    \small 
  	\begin{tabular}{l c c c c c c }
  	\noalign{\hrule height 0.75pt}
         &      &     & \multicolumn{2}{c}{CMN2} & \multicolumn{2}{c}{VAST} \\ \cline{4-7}  
	Sys. & Diar.& DA  & EER	& $C_{\mathrm{prim}}$   &  EER	& $C_{\mathrm{prim}}$   \\ 
	\hline
	1\, \texttt{i} & N & Kaldi& 12.6 & 0.761 & 16.8 & 0.676 \\
    2\, \texttt{x} & N & Kaldi& 11.6 & 0.759 & 15.9 & 0.713 \\
    3\, \texttt{x} & N & Clust. & 8.1 & 0.549 & 14.3 & 0.557 \\
    4\, \texttt{x} & N & Kaldi& 7.5 & 0.452 & {\bf 12.1} & {\bf 0.543} \\
    5\, \texttt{x}+ & N & Kaldi& 7.9 & 0.558 & 15.5 & 0.637 \\
    6\, \texttt{x}+ & Y& CORAL+& {\bf 5.9} & {\bf 0.421} & 12.7 & {\bf 0.543} \\
    7\, \texttt{x} & Y& Kaldi& 7.3 & 0.491 & 14.3 & 0.571 \\
    8\, \texttt{x} & N& Kaldi& 8.1 & 0.551 & 14.6 & 0.601 \\
    9\, \texttt{x} & N& Kaldi& 7.5 & 0.482 & 14.3 & 0.533 \\
    10\,\,\texttt{t} & N& Kaldi & 10.5 & 0.678 & 17.1 & 0.720 \\
    11\,\,\texttt{i} & N& Kaldi & 12.4 & 0.755 & 18.7 & 0.700 \\
    12\,\,\texttt{i} & N& -& 16.4 & 0.814 & 21.3 & 0.788 \\
	\noalign{\hrule height 0.75pt} 
   	\end{tabular}
  	\end{center}
  	\label{table:subsysperf}
  \end{table} 
\renewcommand{\arraystretch}{1.0}

\section{Past Lessons and Future Outlook}
\label{sec:pastandfuture}
The I4U consortium participated in five SREs in the past decade from SRE'08 to SRE'18. In this section, we look into past I4U results (fusion and single best) to derive insights and to have a glimpse into the current and possible future trends. To start with, we give a brief synopsis and highlight the major challenges in the past SREs.

\begin{itemize}
    \item SRE'08, 10, and 12 have in common their evaluation sets drawn from the Mixer corpus, or more precisely, different phases of the Mixer corpus~\cite{Cieri2006,Brandschain2008,Brandschain2010}. One unique feature of the Mixer corpus is that it consists not only conversational telephone speech (CTS) but also conversational and interview style speech recorded over microphone channel. Among others, one major challenge put forward was cross-channel enrollment and test. This is referred to as the \emph{short2-short3} core task in SRE'08, where the enrollment utterances are either telephone or microphone speech, while the test utterances could be telephone, microphone, or interview speech. SRE'10 followed similar setup except that the core task were split into nine \emph{common conditions} (CCs) corresponding to various combinations of channel (telephone, interview, or microphone) and vocal efforts (low, normal, or high). A larger train set was also provided. SRE'12 has a more complicated setup in which the enrollment utterances were derived from previous SRE'08 and SRE'10, while the test utterances were drawn from previously undisclosed subset of the Mixer corpora. The number of CCs was reduced to five. 
    
    \item SRE'16 was derived from the \emph{Call-My-Net} corpus~\cite{Jones2017}. Though the evaluation set is much smaller than that of SRE'12 (few hundreds as opposed to few thousands speakers), SRE'16 posed a new challenge in terms of domain mismatch between the train and evaluation sets. In particular, the train set consists of mainly English speech while the evaluation set was in Tagalog (tgl) and Cantonese (yue). The CMN2 partition of SRE'18 is a continuation of SRE'16 where the same \emph{Call-My-Net} protocol was used to collect speech in Tunisian Arabic~\cite{Jones2017}. The VAST partition of SRE'18 explores a new direction of data collection from online video~\cite{Tracey2018}.   
    
    
    
    
    
\end{itemize}

\begin{table}[t!] 
 \small
  	 \caption{Performance of the primary submissions on the development and evaluation sets.}
    \label{table:performance}
  	\vspace{-3ex}
  	\begin{center}
  		\begin{tabu}{l|c|c|c}
  			\tabucline[1pt]{-}
			\textbf{CMN2}  & EER (\%) 	& Min $C_{primary}$  	& Act $C_{primary}$	\\ \hline
  			Development 		& 4.52	    & 0.277		        & 0.290	            \\
  			Evaluation			& 5.11	    & 0.362		        & 0.368		        \\  			\tabucline[1pt]{-}
			\textbf{VAST}& EER (\%) 	& Min $C_{primary}$  	& Act $C_{primary}$ \\ \hline
   			Development 		& 3.70	    & 0.268		        & 0.300			    \\
   			Evaluation 			& 10.18	    & 0.444		        & 0.550	            \\    			\tabucline[1pt]{-}
       \end{tabu}      
  	\end{center}
  	\label{table:sre18eval}
 \end{table} 
 
\begin{figure} [t!]
  \centerline{\includegraphics[width=1.0\linewidth]{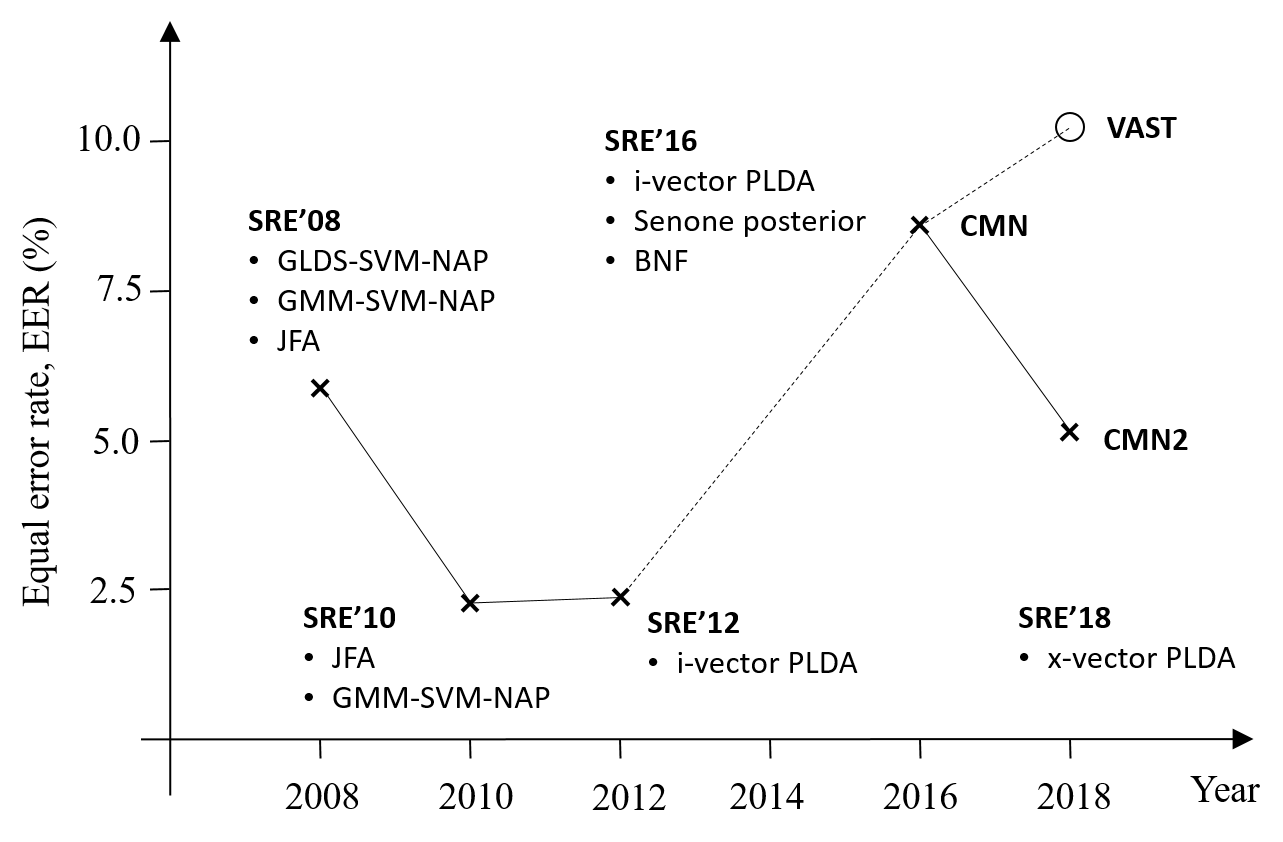}}
  \caption{Progress and performance comparison of I4U submissions from SRE'08 to SRE'18.}
  \label{fig:sres}
  \vspace{-4ex}
\end{figure}

Table \ref{table:i4u_sre} shows the EER of I4U submissions in the past five SREs. Both single-best sub-system and fusion show the same trends. Note that the number of sub-systems used in the fusion varies in each SREs. For SRE'10 and SRE'12, EERs were first computed for each CC and their averages are shown in the table. 
Figure~\ref{fig:sres} shows the evolution of EERs on the evaluation set across five past SREs. 
Strictly speaking, these EERs are not comparable as they were obtained from different evaluation sets. Nevertheless, it is possible to make observations about the general trends. 

\textbf{From SRE'08 to SRE'12}, we see that the EER decreases drastically from SRE'08 at $5.90\%$ to $2.23\%$ in SRE'10 and $2.30\%$ in SRE'12. The main theme in these SREs was channel compensation. In this regard, a larger train set benefited significantly channel compensation techniques like \emph{joint factor analysis} (JFA)~\cite{Kenny2007} and \emph{nuisance attribute projection} (NAP)~\cite{nap} which led to $62\%$ relative EER reduction in SRE'10. In SRE'12, we saw the popularity of i-vector PLDA pipeline~\cite{kenney2010plda} as a simpler alternative to JFA where (i) sequence embedding (i-vector), and (ii) channel compensation and scoring (PLDA) are carried out separately in a pipeline as opposed to a monolithic device. In SRE'12, the EER settled down at similar level as in SRE'10. Compared to its predecessor, the merit of i-vector PLDA is that score normalization is not required.  
Also shown in Figure~\ref{fig:sres} are the GMM-SVM (Gaussian mixture model -- support vector machine)~\cite{nap} and GLDS-SVM (generalized linear discriminant sequence kernel SVM), which were two popular technique that use high-dimensional utterance-level representation with SVM.   

\textbf{From SRE'16 to SRE'18 and beyond}. We witnessed a rebound in EER with the introduction of CMN evaluation set in SRE'16, which posed a different set of challenges compared to SRE'08-12. Language mismatch and lack of labeled in-domain data are among these challenges. In SRE'18, the EER reduces significantly by $51\%$ from $11.48\%$ to $5.58\%$ on SRE'18 CMN2 partition. Undoubtedly, one major contributor is the x-vector deep speaker embedding method~\cite{snyder2018vector,okabe2018attentive}. There is also considerable contribution from unsupervised PLDA adaptation technique as noted in Section~\ref{sec:domain_adaptation}. Another new facet introduced in SRE'18 is the VAST partition. The unconstrained nature of VAST data had proven to be relatively difficult compared to its CMN2 counterpart. We foresee the EER on CMN2 would settle down at around the same level as in SRE'12 when more data is made available. For VAST partition, the difficulty lies at the multi-speaker test segment as noted in Section~\ref{sec:multi-speaker}. In view of the performance gap between the two partitions, we reckon that new breakthrough in speaker diarization aiming at improving speaker recognition accuracy rather than diarization error is necessary. The forthcoming SRE'19 offers another avenue towards that direction with the use of video information~\footnote{https://www.nist.gov/itl/iad/mig/nist-2019-speaker-recognition-evaluation}.          

\textbf{Large-scale fusion} has always been the central stage of I4U submissions. In particular, the I4U submission to SRE'16 encompassed 32 sub-systems, each of them presenting a high-end recognizer involving careful parameter optimization and data engineering. Deploying such massive fusion may be challenging in real use case, reliable fusion indeed plays a key role: it provides a vehicle to solve a common engineering goal, which could not be realistically solved with a single system alone. The SRE'16 fusion result shows that a fairly simple linear fusion improves the performance considerably compared the single-best from 11.48\% to 8.59\% (see Table~\ref{table:i4u_sre}). Interestingly, in the case of SRE'18 CMN2 we do not observe similar large performance gap, indicating the need for new innovations in the underlying technique. Two other useful points that we can derive from I4U experience are: (i) Score pre-calibration before fusion always help. Notably, it allows classifier selection base on their weights. Classifiers with negative correlation with others will have negative weights and could usually be discarded; (ii) Fusion of fusion (\textit{i.e.}, fusing multiple fused systems) is problematic and should be avoided. The rationale is that it tends to over-fit the Development set. 

\textbf{Channel versus domain mismatch.} The notion of channel is used to describe the extrinsic variability imposed on a speech utterance by the acoustic environment, recording device, and the transmission channel. Channel mismatch denotes the inconsistency between the enrollment and test segments in a given trial. For example, a target speaker might be rejected if the channel effects (e.g., enrollment and test utterances of the same speaker but recorded with different devices) is stronger than the speaker characteristic rendered in the utterances. This was the main topic in SRE'08, 10, and 12, and had led to the use of channel compensation techniques, like, JFA~\cite{Kenny2007}, NAP~\cite{nap}, and PLDA~\cite{kenney2010plda}. Domain mismatch, in turn, denotes the inconsistency between Train and Evaluation sets. What this means in the context of SRE'18 CMN2 is that the speaker recognition system was trained with English dataset which is different from from those in which we use the system (i.e., Tunisian Arabic). By domain adaptation, we assume that the channel variability learned from one domain shares some common behaviors in another domain. Simple covariance transformation techniques~\cite{alam2018coral,kalee2018} have shown to work well compared to a much complicated counterpart~\cite{Lin2018}. This is a topic for future research.


\begin{table}[t]
  \small
  \caption{{\it Performance of I4U fusion and single-best submissions in terms of equal-error-rate (EER) on the evaluation set of SRE'08, 10, 12, 16, and 18 \cite{hli2009,hli2010,Saeidi2013,Lee2017,i4u2018}.}}
  \vspace{-4ex}
  \begin{center}
  		\begin{tabu}{l c c c}
  			\tabucline[1pt]{-}
  		                   & \multicolumn{2}{c}{Fusion} & Single best           \\ \cline{2-4}
  			               & \#sub-systems      & EER (\%)    & EER (\%)        \\ \hline
  			SRE'08         & 7                  & 5.90        & 6.10            \\ 
  			SRE'10         & 13                 & 2.23        & 3.55    	    \\		
  			SRE'12         & 17                 & 2.30        & 3.70  	        \\		
   			SRE'16 CMN     & 32                 & 8.59        & 11.48  	        \\		 			
   			SRE'18 CMN2    & 12                 & 5.11        & 5.86  	        \\		 			
   			SRE'18 VAST    & 12                 & 10.18       & 12.06  	        \\	
            \tabucline[1pt]{-}
  		\end{tabu}
  	\end{center}
  	\label{table:i4u_sre}
\end{table}  

\vspace{-2ex}
\section{Conclusions}
This paper presents an overview of the recognition systems and their fusion developed for NIST SRE'18 by I4U consortium. In general, sub-systems that utilized more recent x-vector deep speaker embedding were more successful. On the CMN2 partition, the CORAL+~\cite{kalee2018} unsupervised PLDA adaptation technique has shown to be effective. The VAST partition is more difficult compared to the CMN2. One major challenge is the multi-speaker test segment. Marginal improvement was achieved by pre-processing the multi-speaker test segments with a speaker diarization module. 

Fusion has always been the center stage of I4U submissions. Comparing the single-best and fusion results in the past SREs from SRE'08 to SRE'18, linear fusion optimized with cross-entropy cost works well. We also found that score pre-calibration helps making classifier selection easier. From SRE'08 to SRE'10 and SRE'12, we observed a significant performance gain in I4U submission due to effective channel compensation techniques (joint factor analysis~\cite{Kenny2007} and PLDA~\cite{kenney2010plda}) coupled with a large train set. From SRE'16 to SRE'18, we observed another significant performance gain benefited from the use of deep speaker embedding~\cite{snyder2018vector}.   


\bibliographystyle{IEEEtran}
\bibliography{mybib}
\end{document}